\documentclass[print]{jicspack}

\usepackage{enumerate}
\usepackage{graphicx}
\usepackage{subfigure}
\usepackage{amssymb}
\usepackage[lined,boxed,commentsnumbered, ruled]{algorithm2e}

\begin{document}

\setlength{\parindent}{10pt}

\begin{premaker}


\title{Modeling and Performance Analysis of Pull-Based Live Streaming Schemes in Peer-to-Peer Network}
\author[author1]{Jianwei Zhang},
\author[author1]{Wei Xing\corauthref{cor1}},
\ead{wxing@zju.edu.cn}
\corauth[cor1]{Corresponding author.}
\author[author2]{Yongchao Wang},
\author[author1]{Dongming Lu}
\address[author1]{College of Computer Science and Technology, Zhejiang University, Hangzhou 310027, China}
\address[author2]{Library and Information Center, Zhejiang University, Hangzhou 310027, China}

\begin{abstract}
Recent years mesh-based Peer-to-Peer live streaming has become a promising way for service providers to offer high-quality live video streaming service to Internet users. In this paper, we make a detailed study on modeling and performance analysis of the pull-based P2P streaming systems. We establish the analytical framework for the pull-based streaming schemes in P2P network, give accurate models of the chunk selection and peer selection strategies, and organize them into three categories, i.e., the chunk first scheme, the peer first scheme and the epidemic scheme. Through numerical performance evaluation, the impacts of some important parameters, such as size of neighbor set, reply number, buffer size and so on are investigated. For the peer first and chunk first scheme, we show that the pull-based schemes do not perform as well as the push-based schemes when peers are limited to reply only one request in each time slot. When the reply number increases, the pull-based streaming schemes will reach close to optimal playout probability. As to the pull-based epidemic scheme, we find it has unexpected poor performance, which is significantly different from the push-based epidemic scheme. Therefore we propose a simple, efficient and easily deployed push-pull scheme which can significantly improve the playout probability.
\end{abstract}
\end{premaker}

\section{Introduction}
Recent works for Peer-to-Peer (P2P) streaming can be broadly classified into two categories: multi-tree based streaming and mesh-based streaming [1][2]. The multi-tree based systems organize all peers into one or more multicast trees, and diffuse different substreams along these trees by using application layer multicast routing techniques. Due to the good scalability, low server infrastructure cost and et al., mesh-based streaming has become one of the most important means of P2P live video streaming.

Mesh-based P2P systems can be approximately divided into two categories, \emph{push-based} systems and \emph{pull-based} systems, which are dependent on whether the local peer or the remote peer decides to upload a chunk. In particular, in order to improve the effectiveness of each transmission for a selected chunk, peers in both the push-based and pull-based systems may maintain a list of neighboring peers, and periodically exchange buffer maps with their neighbors.

In push-based systems, there exists the chance that two or more peers try to push the same chunk to the same peer. Usually this problem is already taken into account in most related works. In pull-based (also known as \emph{data-driven}) systems, there exists the chance that two or more peers try to send the same request to the same peer. To address this problem, most literatures assume that the overlay size is large enough to avoid such collisions, or peers have large enough upload bandwidth to deal with all requests from other peers in one time slot. In this paper, we demonstrate this assumption is unreasonable, and confine each peer's ability to process fixed number of requests per time slot. Moreover, we take into account both the random peer selection and random useful peer selection, and classify chunk selection strategies into three categories to facilitate understanding the streaming schemes in pull-based network.

More specifically, the contribution of this paper lies in the following three aspects.
\begin{itemize}
\item We establish the analytical framework for the pull-based streaming schemes in P2P network, and evaluate the peer first scheme, the chunk first scheme and the epidemic scheme in terms of the playout probability and the playout delay. The impacts of some important parameters, such as size of neighbor set, reply number, buffer size and so on are investigated in detail. Our analytical framework and evaluation results reveal the essential distinctions between pull-based streaming schemes and push-based streaming schemes.
\item Unlike previous work towards the pull-based live streaming, we study both the chunk selection and the peer selection strategies analytically. We give accurate formulas of chunk selection strategies, i.e., the latest first selection, the greedy selection and the random selection, and peer selection strategies, i.e., the random selection and random useful selection.
\item More importantly, we also investigate the impact of reply number, i.e., the maximum number of replies a peer can make in one time slot, on all considered streaming schemes.
\item Aiming at enhancing the poor performance of the pull-based epidemic streaming scheme, we propose a simple, efficient and easily deployed push-pull scheme, which needs no buffer map information.
\end{itemize}

The rest of this paper is organized as follows. Section 2 reviews related work. Section 3 describes our analytical framework for pull-based P2P streaming systems. Section 4 gives the analytical models of three categories of streaming schemes. The performance evaluation of all considered streaming schemes is given in Section 5, and we make the conclusion in Section 6.

\section{Related work}
As to the multi-tree based P2P systems, whether or not network coding is adopted has a huge influence on most important conclusions.

Liu et al. [3] theoretically investigate the minimum delay bounds for the single-tree, multi-tree and the proposed snow-ball P2P live streaming algorithm, which can approach the minimum delay bound. They also show the bandwidth heterogeneity among peers can be exploited to significantly improve the delay performance of all peers. Bianchi et al. [4] derive a performance bound for the stream diffusion metric in a homogeneous scenario, and show how this bound relates to the available upload bandwidth and the number of neighbors. Kumar et al. [5]develop a simple stochastic fluid model that accounts for the peers' real-time demand for content, peer churn, bandwidth heterogeneity and playout delay. The model provides closed-form expressions which can be used to shed insight on the fundamental behavior of P2P streaming systems. Liu et al. [6]study the tradeoffs between minimum server load, maximum streaming rate, and minimum tree depth under unconstrained peer selection, single peer selection and constrained peer selection. Through simulations, Magharei et al. [7] show that the inferior performance of the tree-based approach compared with the mesh-based approach should owe to the static mapping of content to a particular overlay tree and diverse placement of peers in different overlay trees.

As to the mesh-based P2P systems, scholars research the push-based systems mostly, but the analytical study of the pull-based systems is very insufficient.

Massoulie et al. [8] give the distributed algorithm that optimally solves the broadcast problem in edge-capacitated networks, and present the decentralized algorithm for solving the node-capacitated broadcast problem, and show some of its optimality properties analytically and through simulation. Sanghavi et al. [9] investigate one-sided push-based and pull-based chunk selection strategies, and propose interleave and advocate two-sided strategies.

Bonald et al. [10] prove that the random peer, latest useful chunk mechanism can achieve dissemination at an optimal rate and within an optimal delay, up to an additive constant term. They use mean-field approximations to derive recursive formulas for the diffusion function of the latest blind chunk, random peer and latest blind chunk, random useful peer schemes. They give simulation results of various practically interesting diffusion schemes in terms of delay, rate, and control overhead. Some schemes can achieve near-optimal performance trade-offs. Fodor et al. [10] extend it by relaxing the complete-graph overlay assumption. They propose an analytical framework that allows the evaluation of scheduling algorithms. They consider the random and nearest strategies, and evaluate the playout delay, playout probability and the scalability of the two strategies. They conclude that the random strategy has fairly good performance for push-based P2P streaming systems. Furthermore, Fodor et al. [12] extend the previous work by considering the effect of the outdated buffer map information, node churn and a larger set of scheduling schemes.

Zhou et al. [13] study two chunk selection strategies, rarest first and greedy, and propose a mixed strategy that combines both of the two strategies. Furthermore, they study the tradeoffs between overlay size, buffer size and playout probability [14]. Zhao et al. [15] present a general and unified mathematical framework to study the large design space of priority-based chunk selection strategies. The analytical framework is asymptotically exact when the overlay size is large. For a given playout probability, the structure of the optimal strategy is fixed as a concatenation of a policy independent of buffer size and the rarest first strategy. Shakkottai et al. [16] study the minimal buffer size needed of the rarest first, greedy and the mixed strategies for a given playout probability, and validate that the mixed strategy can achieve order optimal performance.

Zhang et al. [17] propose an unstructured protocol with a push-pull mechanism to greatly reduce the latency, and evaluate this approach on PlanetLab. Moreover, they [18][19] model scheduling problem in data-driven streaming system as a classical min-cost network flow problem, and then propose both the global optimal scheduling scheme and distributed heuristic algorithm to optimize the system throughput.

Massoulie et al. [20] show the optimality of the random-useful packet forwarding algorithm for edge-capacitated networks, and also show the optimality of the most-deprived neighbor selection scheme combined with random useful packet selection for node-capacitated networks. Abeni et al. [21] presents the formal proof that there exists a distributed scheduling strategy which is able to distribute every chunk to all $N$ peers in exactly   steps. Feng et al. [22] show that there exists a significant performance gap that separates the actual and optimal performance of pull-based mesh protocols. Moreover, periodic buffer map exchanges account for most of this performance gap.

This paper can be thought of as an extension of [13][14] and a complement of [10][12]. Especially, the method of modeling and numerical analysis in [12] is used as reference for our work. In [13][14], the authors assume that the selected peer's uploading bandwidth is large enough to satisfy all the requests in the same time slot. They also simplify the peer selection strategy, i.e., they assume each peer randomly selects another peer to download. In this paper, we relax above assumptions to achieve more important realistic meaning. From another perspective, our work also reveals the difference between the pull-based system and the push-based system studied in [10][12] in both modeling methods and their performance.

\section{System modeling}
As shown in Fig. 1, in a pull-based system neighbor relations between the peers and the forwarding of chunks are determined by local peer $l$. This is the same as in the push-based schemes. However, which peer is chosen to upload is determined by the remote peer $r$. All scheduling decisions are based on the periodic information exchanged between the neighboring peers, which is known as \emph{buffer map}.

Suppose there are one source server and $N$ ordinary peers, and each peer is given $v$ neighbors at random when entering the overlay. In each time slot, the server distributes one chunk to a randomly chosen peer; each peer selects only one chunk and one neighboring peer to request by using the chunk selection strategy and peer selection strategy; each requested peer can upload one or more chunks to the requesting peers. In Fig. 1, the local peer $l$ is willing to send a request for chunk $i$ to one of its $v$ neighbors, the remote peer $r$. Chunk $i$ is available to peer $l$ if peer $l$ does not have this chunk while at least one of its neighbors has this chunk in the playout buffer. Peer $r$ is available to peer $l$ if peer $r$ has at least one available chunk with regard to peer $l$.
\begin{figure}[!h]
\centering
\includegraphics[angle=0, width=0.6\textwidth]{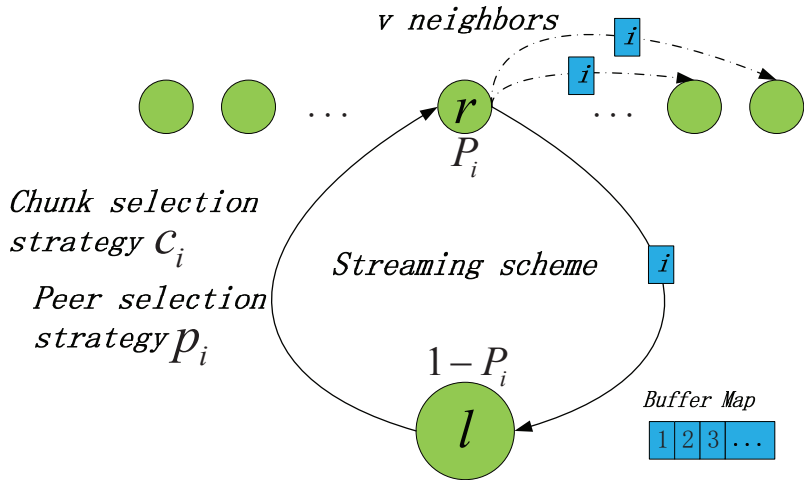}
\caption{System modeling.}
\label{forwarding}
\end{figure}

${{P}_{i}}$ is the probability that a peer is in possession of chunk $i$, or the diffusion rate of the target chunk $i$. The pull-based streaming schemes can be abstracted as the following equation:
\begin{equation}\label{eq:1}
{{P}_{i+1}}={{P}_{i}}+\min \left\{ {{P}_{i}}{{Z}_{i}},1-{{P}_{i}} \right\},{{P}_{1}}=1/N
\end{equation}
where ${{Z}_{i}}$ represents the probability that a peer can successfully obtain the target chunk from one of its neighbors. It is composed of the chunk selection strategy	${{c}_{i}}$ and the peer selection strategy	${{p}_{i}}$. Each peer is given $v$ neighbors at random before the streaming starts.

In order to keep the continuity of the closely related work in Section 2, we make the following assumptions:
\begin{itemize}
\item We assume the buffer status of all peers are independent and follow the same distribution. In other words, all peers' buffer maps are statistically identical, which implies that any chunk in the same position of the buffer has the same presence probability. Furthermore, the same streaming strategy is applied to all peers. This assumption is consistent with [10][11][12][13][14].
\item In the streaming formulas, the neighbor size is considered to be a constant value, although the topology is generated randomly in simulation. This assumption is consistent with [11][12].
\item Peers have homogeneous upload bandwidth, and upload bandwidth rather than download bandwidth is the streaming bottleneck. This assumption holds because in reality users always have asymmetric access bandwidth. This assumption is consistent with [10][11][12][13][14].
\item The impact of node churn will not be covered in this paper. The reason mainly lies in that, by using the approach of Markov process, the authors in [12] argue that the performance of push-based system will not deteriorate in the presence of node churn. Since the streaming schemes in this paper are also based on the mesh topology, whose properties related to node churn are similar to the push-based schemes in [12], we speculate that our schemes have good robustness as well.
\end{itemize}

\section{Live streaming schemes}
\subsection{The transmission of a single chunk}
In order to understand the pull-based scheme more intuitively, we will examine two special cases that there is only one chunk to transmit, namely ${{c}_{i}}=1$, this leads to the random useful peer selection and the random peer selection.

\subsubsection{Single chunk, random useful peer selection}

In fact, this is the simplest case, we have ${{p}_{i}}=1$. Equation (1) can be simplified to:
\begin{equation}\label{eq:2}
{{Z}_{i}}=1-{{P}_{i}}^{v}
\end{equation}

We can derive a closed-form solution of the distribution ${{P}_{i}}$ from the difference equation for ${{P}_{i}}$. We can convert the discrete difference equation into the following continuous differential equation:
\begin{equation}\label{eq:3}
y'=y(1-{{y}^{v}})
\end{equation}

The symbol y stands for ${{P}_{i}}$ and the symbol x corresponds to $i$ in the discrete case. These continuous differential equations can be derived by substituting $\frac{dy}{dx}$ for $\frac{{{P}_{i+1}}-{{P}_{i}}}{(i+1)-i}$ and $y$ for ${{P}_{i}}$. Solving above equation, we obtain the following equation:
\begin{equation}\label{eq:4}
y=1-{{({{e}^{vx+C}}+1)}^{-1}},C=\ln \left[ {{(1-{{P}_{1}})}^{-1}}-1 \right]-v
\end{equation}

\subsubsection{Single chunk, random peer selection}

In this case, we have ${{p}_{i}}=\frac{1}{v}$. Equation(1) can be simplified to:
\begin{equation}\label{eq:5}
{{Z}_{i}}=(1-{{P}_{i}})\left[ 1-{{(1-{{p}_{i}})}^{v}} \right],{{p}_{i}}=\frac{1}{v}
\end{equation}
\begin{equation}\label{eq:6}
{{P}_{i+1}}={{P}_{i}}+{{P}_{i}}(1-{{P}_{i}})(1-{{e}^{-1}})
\end{equation}

Solving above equation in the similar method, we obtain:
\begin{equation}\label{eq:7}
x={{(1-{{e}^{-1}})}^{-1}}\left[ \ln y+\ln (1-y) \right]+C,C=1-{{(1-{{e}^{-1}})}^{-1}}\left[ \ln {{P}_{1}}+\ln (1-{{P}_{1}}) \right]
\end{equation}

\subsection{Streaming of multiple chunks}
\subsubsection{Chunk selection modes and chunk selection strategies}

In pull-based P2P streaming systems, chunk $k$ is available if at least one of its neighbors has this chunk while the local peer not. The probability that chunk $k$ is available is:
\begin{equation}\label{eq:8}
{{Q}_{v,k}}=(1-{{P}_{k}})\left[ 1-{{(1-{{P}_{k}})}^{v}} \right]
\end{equation}

According to how many neighbors' buffer map knowledge should be obtained when making the chunk selection decision, we define the following three \emph{chunk selection modes} ${{c}_{0,i}}$, ${{c}_{1,i}}$ and ${{c}_{v,i}}$, and all of them are integrated into ${{c}_{i}}$ formally for expression convenience.

\begin{itemize}
\item \emph{0-neighbor chunk selection}: ${{c}_{0,i}}$ is the chunk selection strategy only based on its own buffer, no other buffer map knowledge is needed.
\item \emph{1-neighbor chunk selection}: ${{c}_{1,i}}$ is the chunk selection strategy based on one of its neighbors' buffer map. In this mode, it is assumed that the neighbor is already chosen before chunk selection.
\item \emph{$v$-neighbor chunk selection}: ${{c}_{v,i}}$ is the chunk selection strategy based on all its $v$ neighbors' buffer maps.
\end{itemize}

Although ${{c}_{1,i}}$ can be regarded as to a special form of ${{c}_{v,i}}$ when $v=1$, their difference lies in that ${{c}_{1,i}}$ is used for peer first schemes and ${{c}_{v,i}}$ is used for chunk first schemes.

Noticeably, each chunk selection mode contains all three chunk selection strategies, i.e., \emph{latest first chunk selection}, \emph{greedy chunk selection} and \emph{random chunk selection}.

1) Latest first chunk selection

For the latest chunk selection, a peer will select a chunk which has the fewest number of copies in the swarm. In P2P file sharing systems, the latest chunk selection is also known as the rarest first chunk selection. Under this strategy, the selection priority is given from chunk 1 to chunk $n$ in its buffer. Therefore the latest chunk selection strategy can be expressed as follows:
\begin{equation}\label{eq:9}
{{c}_{0,i}}=\prod\limits_{k=1}^{i-1}{{{P}_{k}}}
\end{equation}
\begin{equation}\label{eq:10}
{{c}_{1,i}}=\prod\limits_{k=1}^{i-1}{(1-{{Q}_{1,k}})}
\end{equation}
\begin{equation}\label{eq:11}
{{c}_{v,i}}=\prod\limits_{k=1}^{i-1}{(1-{{Q}_{v,k}})}
\end{equation}

The first one is called \emph{latest blind chunk selection} in many literatures, and the other two are called \emph{latest useful chunk selection}.

2) Greedy chunk selection

For the greedy chunk selection, a peer will select a chunk which is nearest to its playout point. Under this strategy, the selection priority is given from chunk $n$ to chunk 1 in its buffer. In contrast to the latest chunk selection, the nearest chunk selection strategy can be expressed as:
\begin{equation}\label{eq:12}
{{c}_{0,i}}=\prod\limits_{k=i+1}^{n-1}{{{P}_{k}}}
\end{equation}
\begin{equation}\label{eq:13}
{{c}_{1,i}}=\prod\limits_{k=i+1}^{n-1}{(1-{{Q}_{1,k}})}
\end{equation}
\begin{equation}\label{eq:14}
{{c}_{v,i}}=\prod\limits_{k=i+1}^{n-1}{(1-{{Q}_{v,k}})}
\end{equation}

3) Random chunk selection

For the random chunk selection, a peer will select a chunk in its buffer randomly. Suppose ${{A}_{i}}$ is all possible chunk sets exclusive of chunk $i$ with size $\left| {{A}_{i}} \right|$, and $\overline{{{A}_{i}}}$ is the complementary chunk sets of ${{A}_{i}}$, which still excludes chunk $i$. Thus we have $\left| {{A}_{i}} \right|+\left| \overline{{{A}_{i}}} \right|=n-1$.
\begin{equation}\label{eq:15}
{{c}_{0,i}}=\sum\limits_{\left| {{A}_{i}} \right|=0}^{n-1}{\left\{ \frac{1}{\left| {{A}_{i}} \right|+1}\sum\limits_{\forall {{A}_{i}}}{\left[ \prod\limits_{k\in {{A}_{i}}}{(1-{{P}_{k}})}\prod\limits_{k\in {{{\bar{A}}}_{i}}}{{{P}_{k}}} \right]} \right\}}
\end{equation}
\begin{equation}\label{eq:16}
{{c}_{1,i}}=\sum\limits_{\left| {{A}_{i}} \right|=0}^{n-1}{\left\{ \frac{1}{\left| {{A}_{i}} \right|+1}\sum\limits_{\forall {{A}_{i}}}{\left[ \prod\limits_{k\in {{A}_{i}}}{{{Q}_{1,k}}}\prod\limits_{k\in {{{\bar{A}}}_{i}}}{(1-{{Q}_{1,k}})} \right]} \right\}}
\end{equation}
\begin{equation}\label{eq:17}
{{c}_{v,i}}=\sum\limits_{\left| {{A}_{i}} \right|=0}^{n-1}{\left\{ \frac{1}{\left| {{A}_{i}} \right|+1}\sum\limits_{\forall {{A}_{i}}}{\left[ \prod\limits_{k\in {{A}_{i}}}{{{Q}_{v,k}}}\prod\limits_{k\in {{{\bar{A}}}_{i}}}{(1-{{Q}_{v,k}})} \right]} \right\}}
\end{equation}

\subsubsection{Different streaming schemes}

According to three chunk selection modes introduced above, we divide the pull-based streaming schemes into three categories: \emph{chunk first} (CF) scheme, \emph{peer first} (PF) scheme and \emph{epidemic} (EP) scheme.

1) Chunk first scheme

First, the local peer chooses a target chunk by using the $v$-neighbor chunk selection mode according to all its $v$ neighbors buffer maps and its own buffer status. Then, the peer selects uniformly at random among its neighbors that is in possession of the selected chunk.

2) Peer first scheme

First, the local peer selects a neighbor from the set of available neighbors randomly based on their buffer maps and its own buffer status. Then, the peer selects a target chunk according to the 1-neighbor chunk selection mode. When $v=1$, the peer first scheme is equivalent to the chunk first scheme.

3) Epidemic scheme

First, the local peer chooses a target chunk by using the 0-neighbor chunk selection mode just according to its own buffer status. Then, the peer selects a neighbor uniformly at random, whether or not it has the selected chunk.

The main characteristic of the epidemic scheme is to diffuse chunks efficiently without knowing neighboring peers' buffer maps. Unfortunately, as can be observed in Section 5.1, the pull-based epidemic scheme has very poor performance.

Considering the complicated relationship existent in the chunk selection and the peer selection strategies, we make the summary for our analytical framework for the pull-based P2P streaming schemes in Fig. 2. Obviously, this will lead to a lot of combinations.
\begin{figure}[!h]
\centering
\includegraphics[angle=0, width=0.7\textwidth]{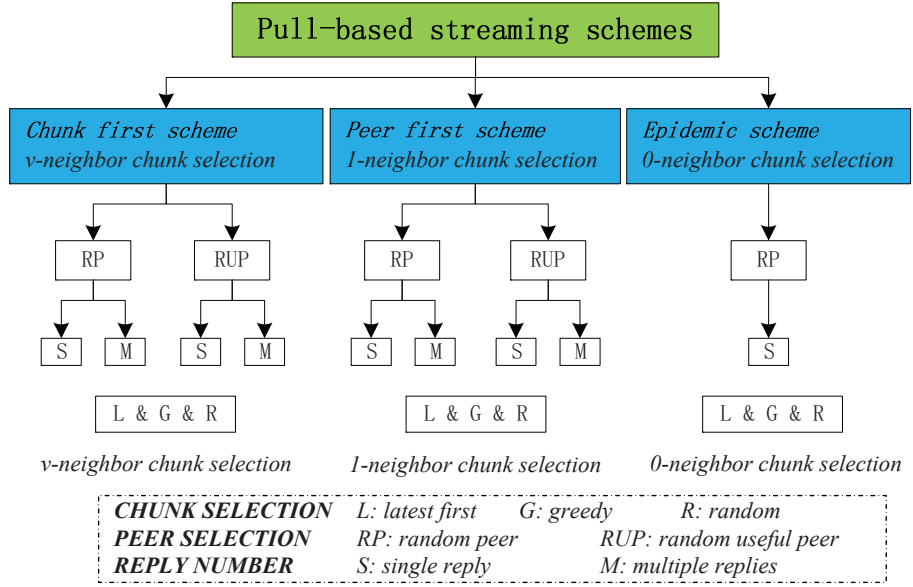}
\caption{Analytical framework for the pull-based P2P streaming schemes.}
\label{forwarding}
\end{figure}

\subsubsection{Analytical models}

In accordance with the four streaming schemes in Fig. 2, we derive the analytical models correspondingly in this subsection.

1) Chunk first scheme

The probability that a peer will receive a request for chunk $i$ from one of its $v$ neighbors is denoted by ${{r}_{i}}$, then we have:
\begin{equation}\label{eq:18}
{{r}_{i}}={{p}_{i}}{{c}_{v,i}}(1-{{P}_{i}})
\end{equation}

The probability that a peer will receive at least one such request is $1-{{\left( 1-{{r}_{i}} \right)}^{v}}$. If a peer is restricted to reply only one request in one time slot, which is referred to as \emph{single reply} hereinafter, then we have the following equation:
\begin{equation}\label{eq:19}
{{Z}_{i}}=1-{{\left( 1-{{r}_{i}} \right)}^{v}}
\end{equation}

Suppose a peer can simultaneously send $U$ (\emph{reply number}) chunks in one time slot, which is referred to as \emph{multiple replies} hereinafter. If it has received $k$ requests, it can reply to at most ${{s}_{k}}$ requesting peers in one time slot, therefore we have:
\begin{equation}\label{eq:20}
{{Z}_{i}}=\sum\limits_{k=1}^{v}{{{s}_{k}}C_{v}^{k}{{r}_{i}}^{k}{{(1-{{r}_{i}})}^{v-k}}},{{s}_{k}}=\min \{k,U\}
\end{equation}

In the chunk first scheme, the peer selection falls into random peer selection and random useful peer selection.

a) Random peer selection

In this case, the peer selection strategy can be simply expressed as:
\begin{equation}\label{eq:21}
{{p}_{i}}=\frac{1}{v}
\end{equation}

When $v$ is large enough, equation (19) can be approximated by the following equation.
\begin{equation}\label{eq:22}
{{Z}_{i}}=1-{{e}^{-{{c}_{v,i}}(1-{{P}_{i}})}}
\end{equation}

b) Random useful peer selection

In this case, ${{p}_{i}}$ is the probability that one of a peer's neighbors who possess chunk $i$ is chosen to send request to.
\begin{equation}\label{eq:23}
{{p}_{i}}=\sum\limits_{k=0}^{v-1}{\frac{1}{k+1}C_{v-1}^{k}{{P}_{i}}^{k}{{(1-{{P}_{i}})}^{v-1-k}}}
\end{equation}

2) Peer first scheme

When peer selection is executed first, the probability that a peer is chosen at least once by all its neighbors is $1-{{(1-{{p}_{i}})}^{v}}$. If a peer can only send a single reply in one time slot, the probability that chunk $i$ is chosen to request by the requesting peer is ${{c}_{1,i}}(1-{{P}_{i}})$, then we have:
\begin{equation}\label{eq:24}
{{Z}_{i}}={{c}_{1,i}}(1-{{P}_{i}})\left[ 1-{{(1-{{p}_{i}})}^{v}} \right]
\end{equation}

In the case of multiple replies, denote that ${{w}_{i}}={{c}_{1,i}}(1-{{P}_{i}})$, it represents the probability that a peer will choose chunk $i$. Suppose a peer with upload bandwidth $U$ has received $k$ requests, then it can reply to at most ${{s}_{k}}$ requesting peers in one time slot. ${{q}_{k}}$ calculates the expectation of the number of peers that a peer can reply to. Therefore we have:
\begin{equation}\label{eq:25}
{{q}_{k}}=\sum\limits_{t=1}^{{{s}_{k}}}{tC_{{{s}_{k}}}^{t}{{w}_{i}}^{t}{{(1-{{w}_{i}})}^{{{s}_{k}}-t}}},{{s}_{k}}=\min \{k,U\}
\end{equation}
\begin{equation}\label{eq:26}
{{Z}_{i}}=\sum\limits_{k=1}^{v}{{{q}_{k}}C_{v}^{k}{{p}_{i}}^{k}{{(1-{{p}_{i}})}^{v-k}}}
\end{equation}

The peer first scheme can be also divided into random peer selection and random useful peer selection.

a) Random peer selection

In this case, whether a neighboring peer is selected before chunk selection or not makes no difference to the whole scheme. Therefore we have:
\begin{equation}\label{eq:27}
{{p}_{i}}=\frac{1}{v}
\end{equation}
\begin{equation}\label{eq:28}
{{Z}_{i}}={{c}_{1,i}}(1-{{P}_{i}})(1-{{e}^{-1}})
\end{equation}

Since ${{q}_{k}}=k{{w}_{i}}$ in this case, it can be simplified to the following equation which is just the core relation in [13]:
\begin{equation}\label{eq:29}
{{Z}_{i}}={{c}_{1,i}}(1-{{P}_{i}})
\end{equation}

b) Random useful peer selection

${{u}_{i}}$ is the probability that the target peer is useful to the local peer l, then:
\begin{equation}\label{eq:30}
{{u}_{i}}=1-\prod\limits_{k=1}^{n}{(1-{{Q}_{1,k}})}
\end{equation}

Then, the probability that peer $l$ selects peer $r$ is:
\begin{equation}\label{eq:31}
{{p}_{i}}=\sum\limits_{k=0}^{v-1}{\frac{1}{k+1}C_{v-1}^{k}{{u}_{i}}^{k}{{(1-{{u}_{i}})}^{v-1-k}}}
\end{equation}

3) Epidemic scheme

In this scheme, we have ${{p}_{i}}=\frac{1}{v}$, and the probability that chunk $i$ is chosen to request is ${{c}_{0,i}}(1-{{P}_{i}})$, then we have:
\begin{equation}\label{eq:32}
{{Z}_{i}}={{c}_{0,i}}(1-{{P}_{i}})\left[ 1-{{(1-{{p}_{i}})}^{v}} \right],{{p}_{i}}=\frac{1}{v}
\end{equation}

When $v$ is large, it can be simplified to:
\begin{equation}\label{eq:33}
{{Z}_{i}}={{c}_{0,i}}(1-{{P}_{i}})(1-{{e}^{-1}})
\end{equation}

If peers can send any number of chunks at a time, the constant factor $1-{{e}^{-1}}$ in equation (33) is removed. This will offer no substantial performance advantage than the single reply case. In consequence, although the equation of multiple replies can be obtained similar to other two schemes, we think it is unnecessary.
\section{Performance evaluation}
Due to the high complexity of the difference equations in Section 4.2, we can't obtain their analytical solutions as in Section 4.1. As a consequence, we will solve the above numerically in an iterative way. We implement a discrete event-driven simulator to examine our models. In the following simulations, unless otherwise noted, the overlay size, buffer size and the size of neighbor set is set to $N=100$, $n=40$ and $v=10$ respectively. On the other hand, although we can designate the same neighbor set size $v$ as in the models by introducing a regular graph in simulation, e.g., a circumference-like topology, we prefer to generate a random graph to make the simulation more realistic. In our topology, there may be a few peers connecting to less than $v$ neighbors, and we think this will partly account for the discrepancy between our models and simulation results.
\subsection{Model validation}
The buffer status is somewhat similar to a snapshot during the streaming process. In Fig. 3, we can see that compared with the push-based streaming schemes investigated in [12], not only the growth rate of diffusion but the playout probability are limited seriously. On the other hand, the peer first scheme considerably outperforms other two schemes when the latest first strategy and the greedy strategy are used, while this is not the truth when random strategy is taken into use. Noticeably, the performance of the epidemic scheme is very poor, which is totally unexpected.

All numerical results of our models fit the simulation results very well, with the only exception that the random strategy in the chunk first scheme seems to overestimate. We think the reason is that, compared to our models, more chunks are concentrated in the same peer in simulation, while peers can send only one chunk at a time. This will be serious when random chunk selection strategy is employed. As can be seen in Section 5.2.2, with the increasing of the reply number $U$, the simulation performance of random strategy grows quickly at the same level as the model.
\begin{figure}[!h]
\centering
\subfigure[]{
\label{fig:subfig:a}
\includegraphics[width=0.32\textwidth]{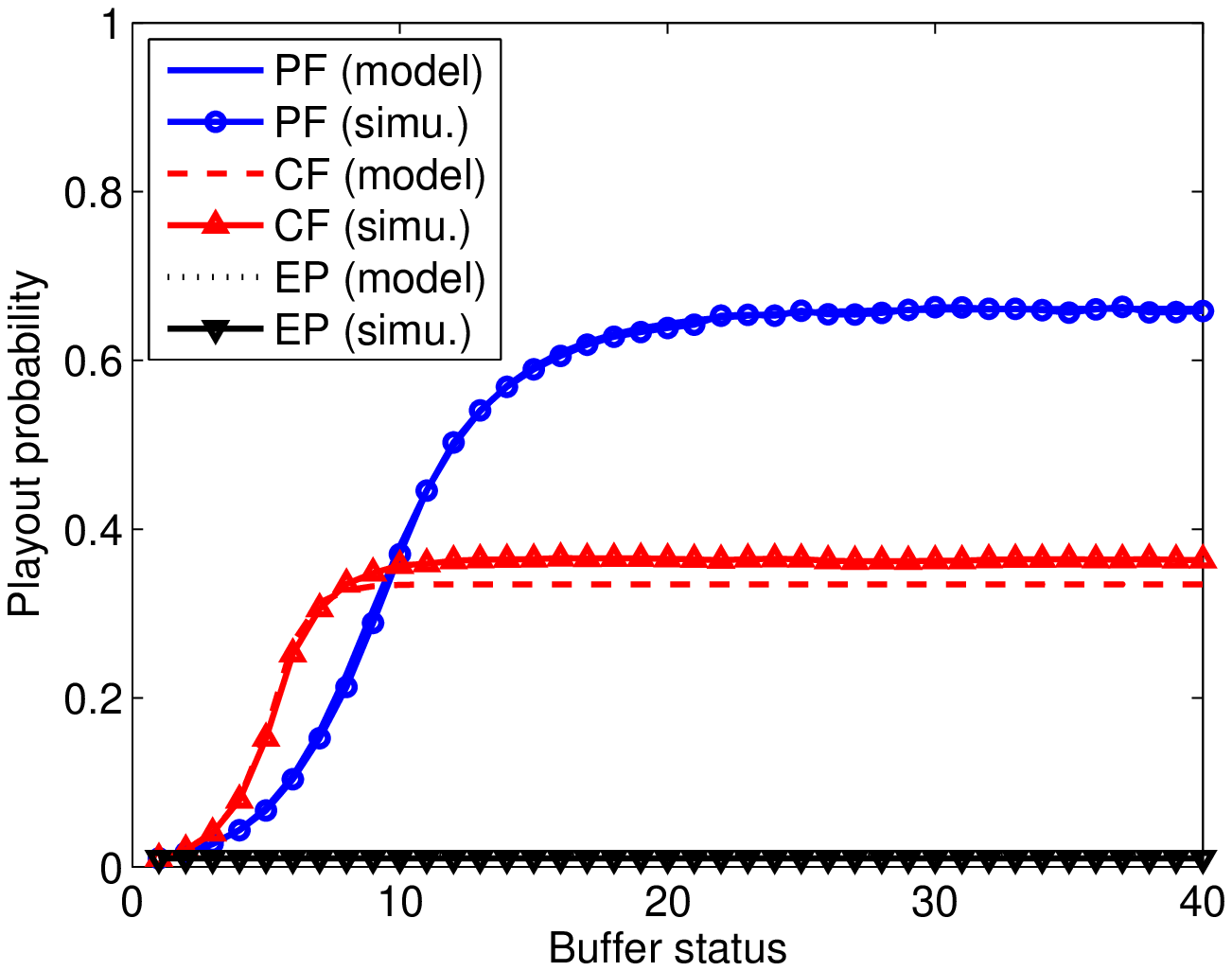}}
\subfigure[]{
\label{fig:subfig:b}
\includegraphics[width=0.32\textwidth]{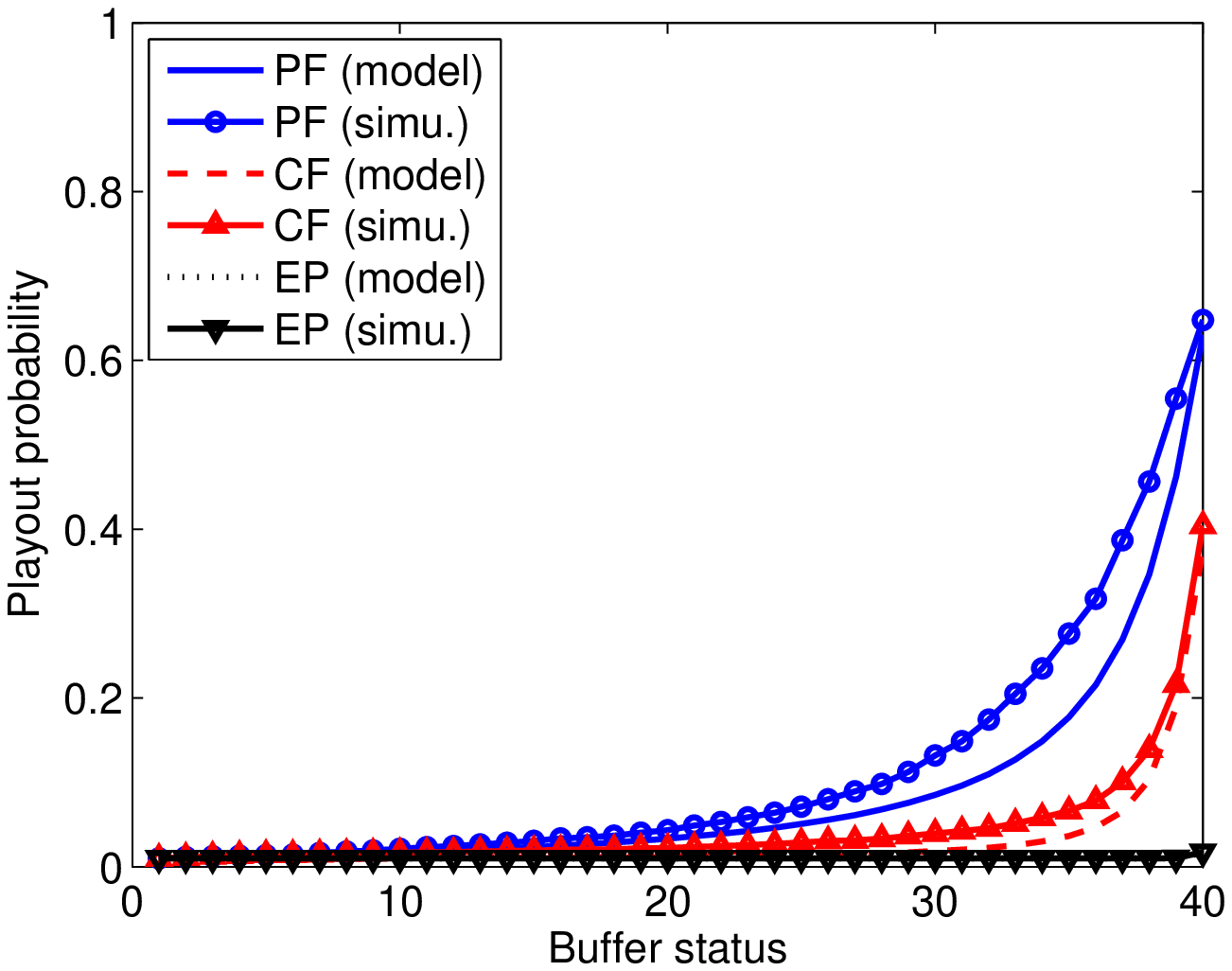}}
\subfigure[]{
\label{fig:subfig:c}
\includegraphics[width=0.32\textwidth]{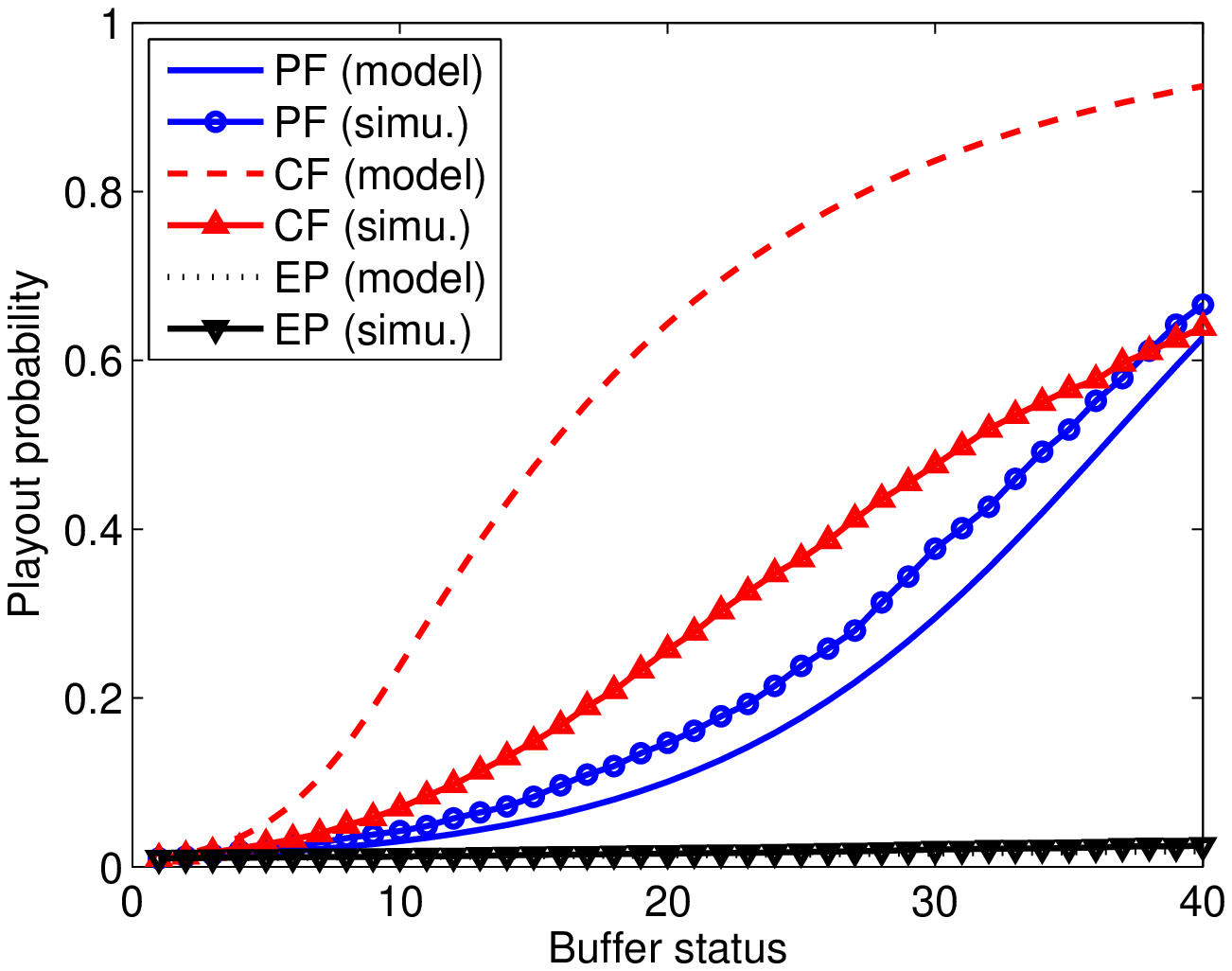}}
\caption{Buffer status of three schemes, $U=1$.(a) Latest first; (b) Greedy; (c) Random.}
\label{fig:subfig}
\end{figure}
\subsection{Peer first scheme and chunk first scheme}
\subsubsection{Impact of playout delay and neighbor set size}

In this subsection, we will investigate the performance of the peer first and chunk first schemes in terms of the playout probability.

Fig. 4(a) illustrates the playout probability versus the playout delay for all chunk first and peer first schemes. When playout delay reaches some value, the playout probability will converge to a stable value. In chunk first scheme, the random strategy outperforms other two chunk selection strategies by a large margin. While in peer first scheme, the latest first strategy is the best, and the playout probability of three chunk selection strategies become close when playout delay grows.

Fig. 4(b) shows that, the playout probability of the chunk first schemes will decreases rapidly as the size of neighbor set increases, with the exception of the random strategy. This is quite different from the push-based streaming schemes mentioned in [10], whose playout probability remains the same with the increasing of the size of neighbor set. As to the peer first schemes, the playout probability will drop slowly and remain above 0.6 when the size of neighbor set grows. In reality, on account of the limitation of the topology connectivity, the performance may be very poor when the neighbor set size becomes too small.
\begin{figure}[!h]
\centering
\subfigure[]{
\label{fig:subfig:a}
\includegraphics[width=0.4\textwidth]{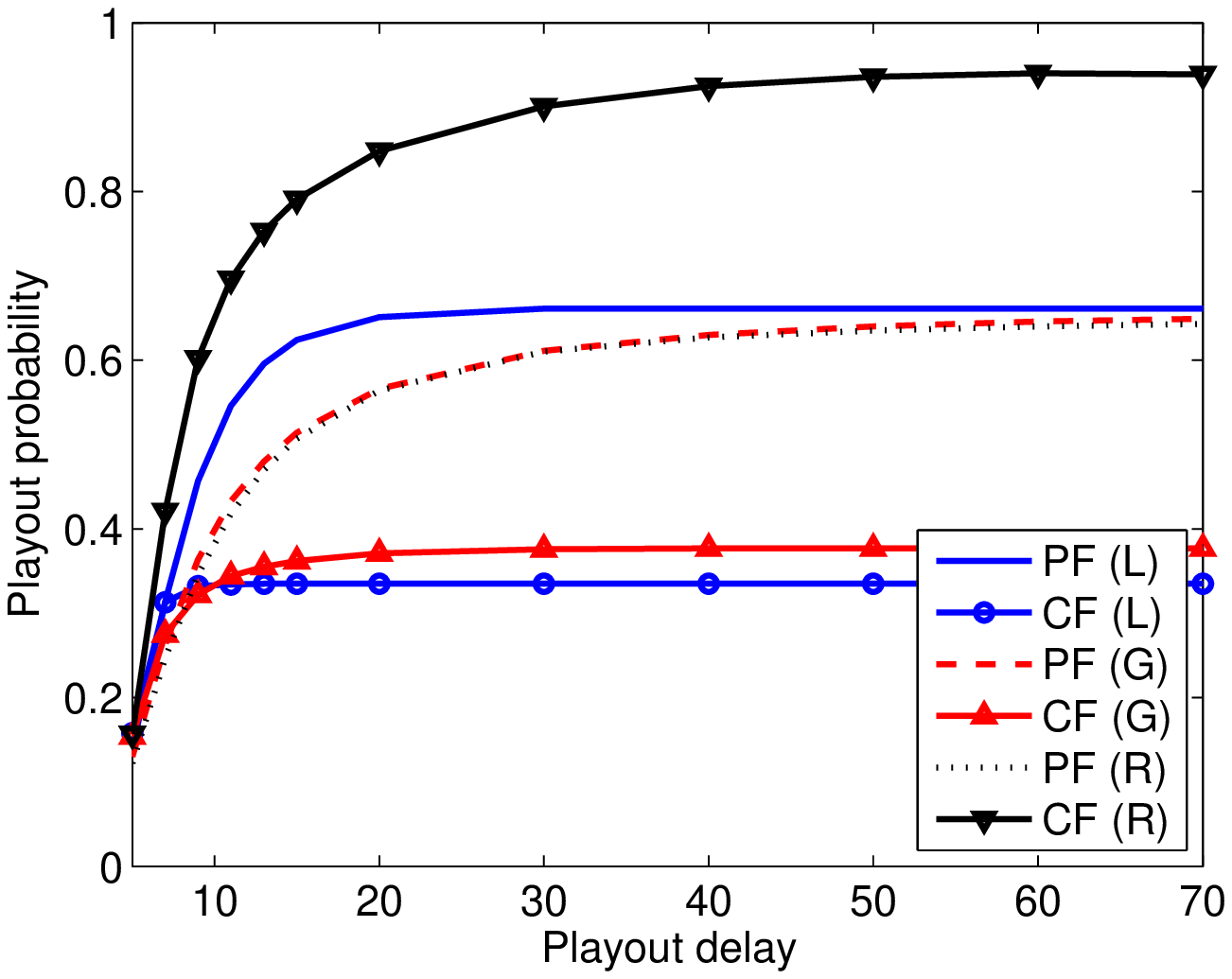}}
\subfigure[]{
\label{fig:subfig:b}
\includegraphics[width=0.4\textwidth]{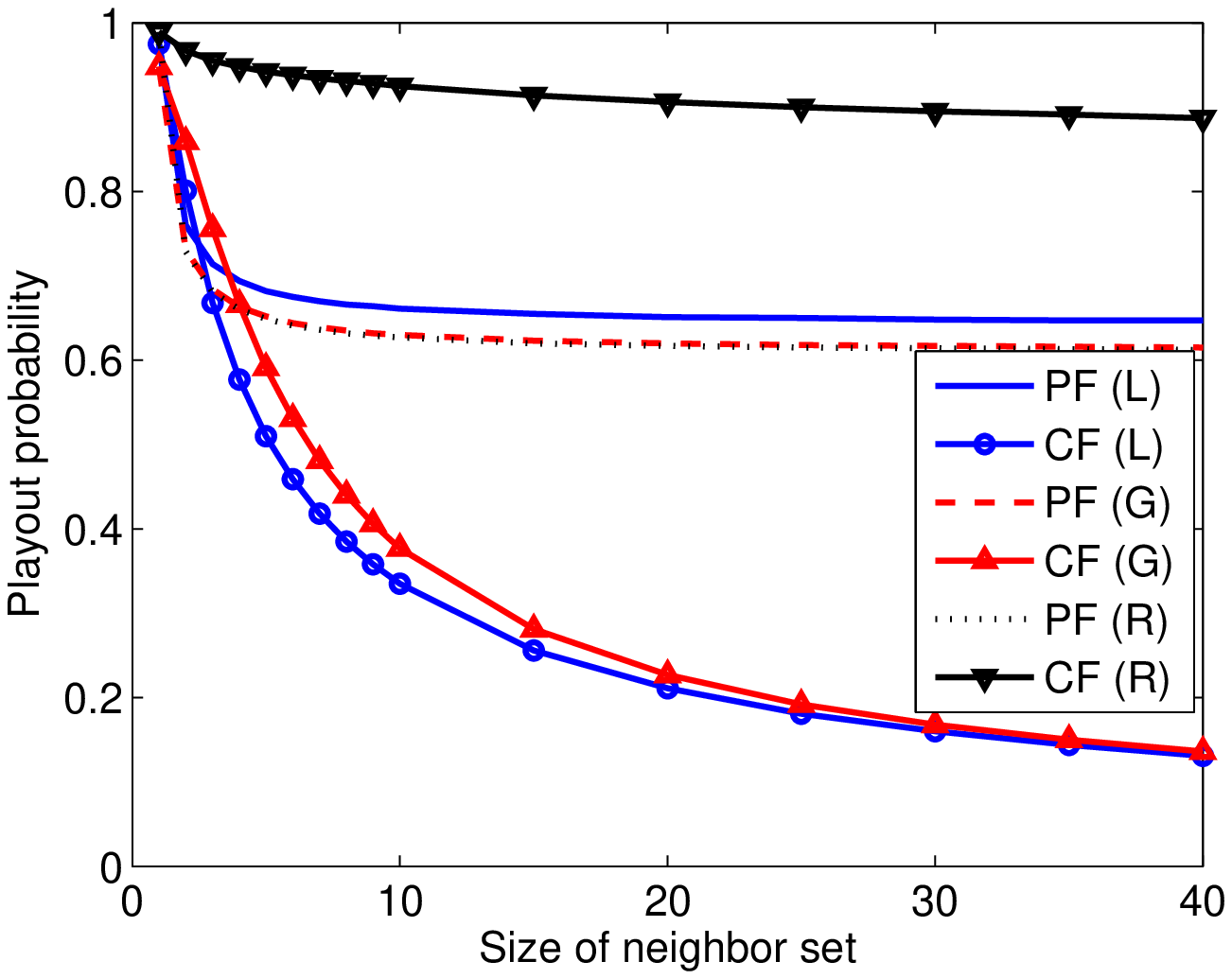}}
\caption{(a) Playout probability vs.Playout delay; (b) Playout probability vs. Size of neighbor set.}
\label{fig:pc}
\end{figure}

\subsubsection{Multiple replies}

Fig. 5 plots the buffer status when $U=4$, we can see our models fit the simulation results very well. All three chunk selection strategies in the peer first scheme become very close, and reach the optimal playout probability. In the chunk first scheme, the random strategy gets a good trade-off between a faster growth rate and a higher playout probability.
\begin{figure}[!h]
\centering
\subfigure[]{
\label{fig:subfig:a}
\includegraphics[width=0.32\textwidth]{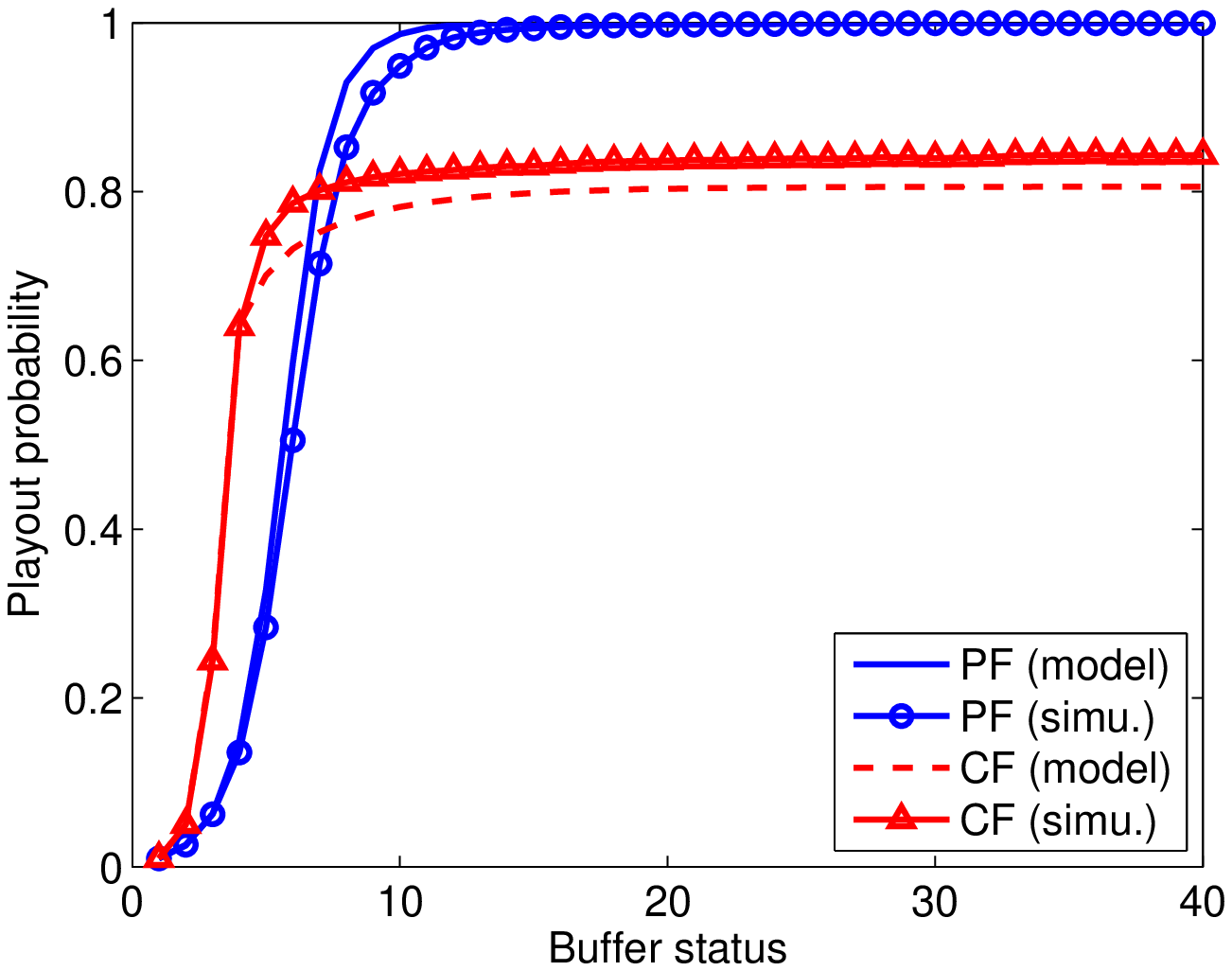}}
\subfigure[]{
\label{fig:subfig:b}
\includegraphics[width=0.32\textwidth]{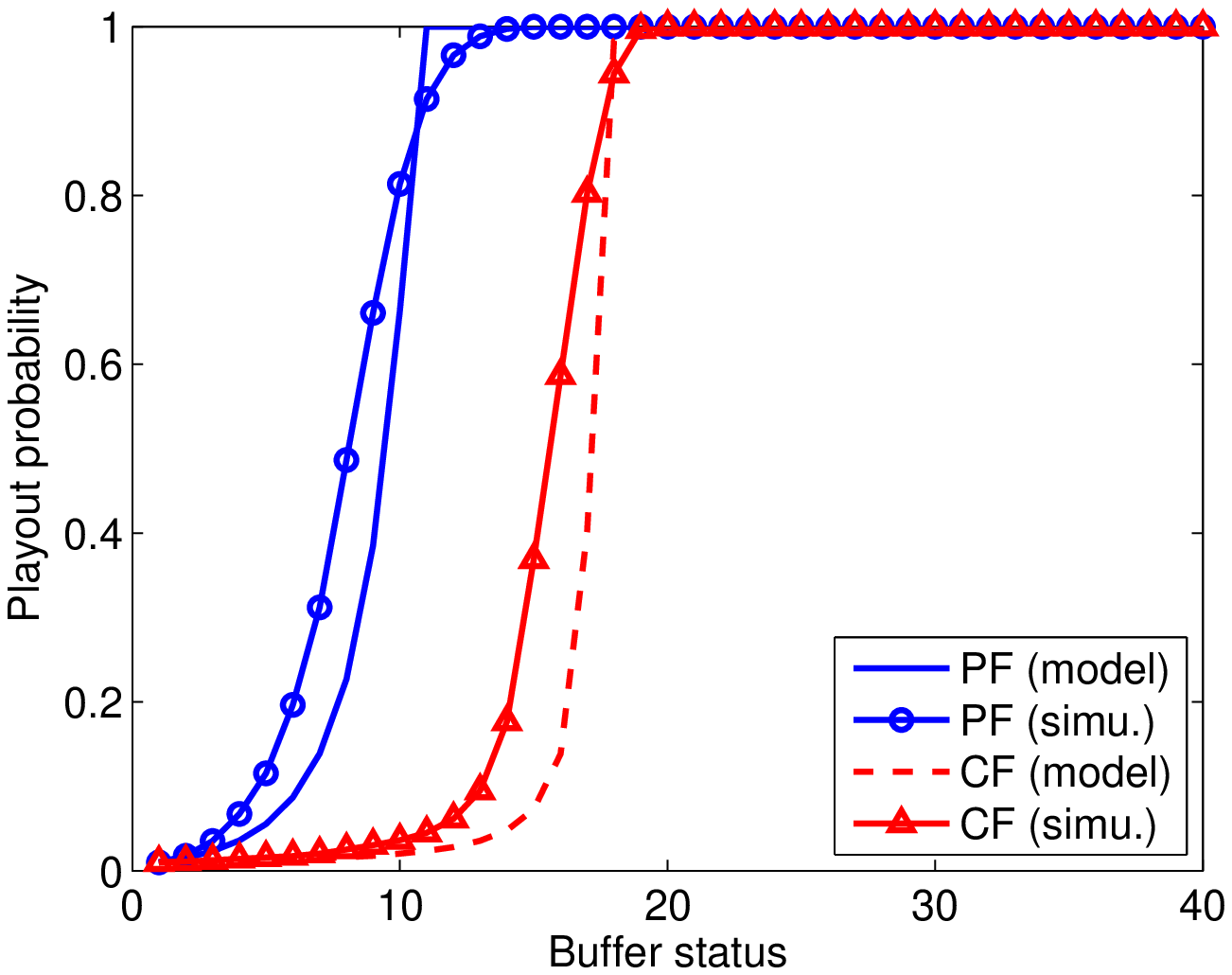}}
\subfigure[]{
\label{fig:subfig:c}
\includegraphics[width=0.32\textwidth]{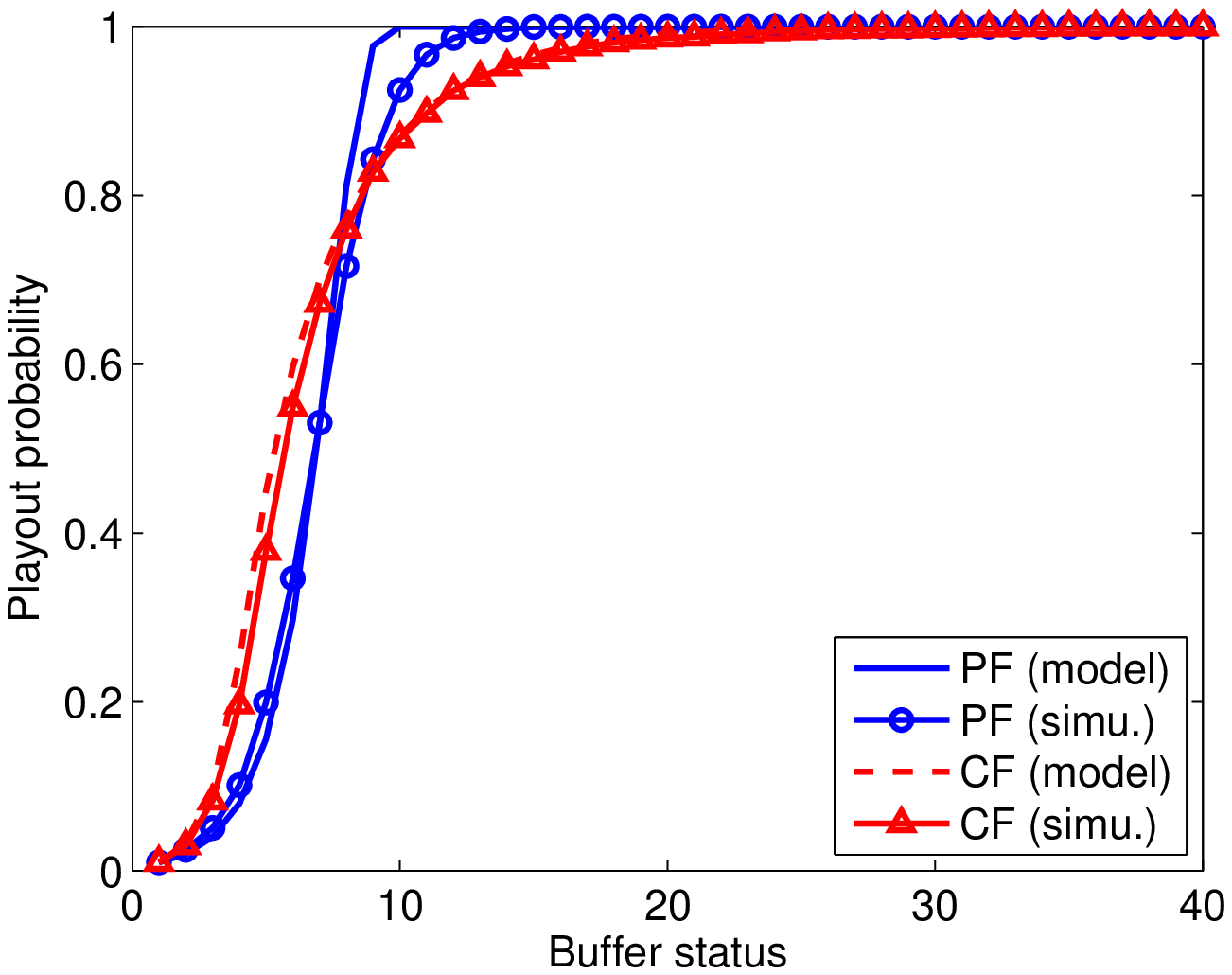}}
\caption{Buffer status of PF scheme and CF scheme, $U=4$. (a) Latest first; (b) Greedy; (c) Random.}
\label{fig:subfig}
\end{figure}

Fig. 6 illustrates the variation of playout probability under different reply number $U$. When $U$ is larger than 3, all strategies except the latest first strategy in the chunk first scheme will be asymptotically optimal.
\begin{figure}[!h]
\centering
\includegraphics[angle=0, width=0.4\textwidth]{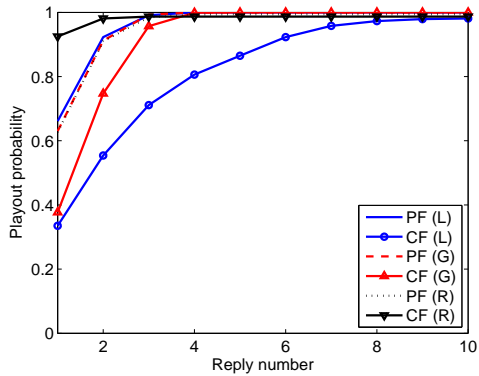}
\caption{Playout probability vs. Replay number $U$ under three chunk selection strategies.}
\label{forwarding}
\end{figure}
\subsection{Epidemic scheme}
The main characteristic of the epidemic scheme is to diffuse chunks efficiently without knowing neighboring peers' buffer maps. Due to the poor performance of the epidemic scheme, we propose the \emph{push-pull scheme}, which still belongs to the epidemic scheme in essence.

We split the buffer into two partitions by a fixed position $d$, $d\in [1,n]$. In each time slot, each peer sends a request to one of its neighbors at random in sequence from n to $d+1$, namely the greedy chunk selection strategy is used. On the other hand, each peer chooses a chunk it already holds in greedy order to send to a peer who requests this chunk at random. If a peer cannot reply a request in $[d+1,n]$, it executes the latest-first push to a randomly chosen neighbor. If no chunk in $[d+1,n]$ is available to reply to the requesting peer, which can be calculated by $1-{{P}_{d+1}}$, the push scheme which employs the latest first chunk selection strategy is performed from buffer position 1 to $d$. Therefore we have the following segmented equation.
\begin{equation}\label{eq:34}
{{P}_{i+1}}=\left\{ \begin{array}{l}
  {{P}_{i}}+(1-{{P}_{i}})(1-{{e}^{-{{P}_{i}}{{c}_{push,i}}}}),{{c}_{push,i}}=(1-{{P}_{d+1}})\prod\limits_{k=1}^{i-1}{(1-{{P}_{k}})},i\in [1,d] \\
 {{P}_{i}}+{{P}_{i}}{{c}_{0,i}}(1-{{P}_{i}})(1-{{e}^{-1}}),i\in [d+1,n-1] \\
\end{array} \right.
\end{equation}

When $d=n$, the push-pull scheme is the same as the push scheme in [10]; when $d=1$, the push-pull scheme becomes the same as the pull scheme . For given buffer size and overlay size, we could get an optimal $d$ by a simple traversal search method. This will not involve a large amount of calculation, as the buffer size is very limited in practice. Here we only give a numerical result to show the playout probability of our push-pull scheme under different size of d in Fig. 7(a).

Fig. 7(b) validates the advantage of our push-pull scheme over the traditional push scheme in [10] when $d=20$. Compared to the push scheme, the playout probability of the push-pull scheme is raised by around 20\%.
\begin{figure}[!h]
\centering
\subfigure[]{
\label{fig:subfig:a}
\includegraphics[width=0.4\textwidth]{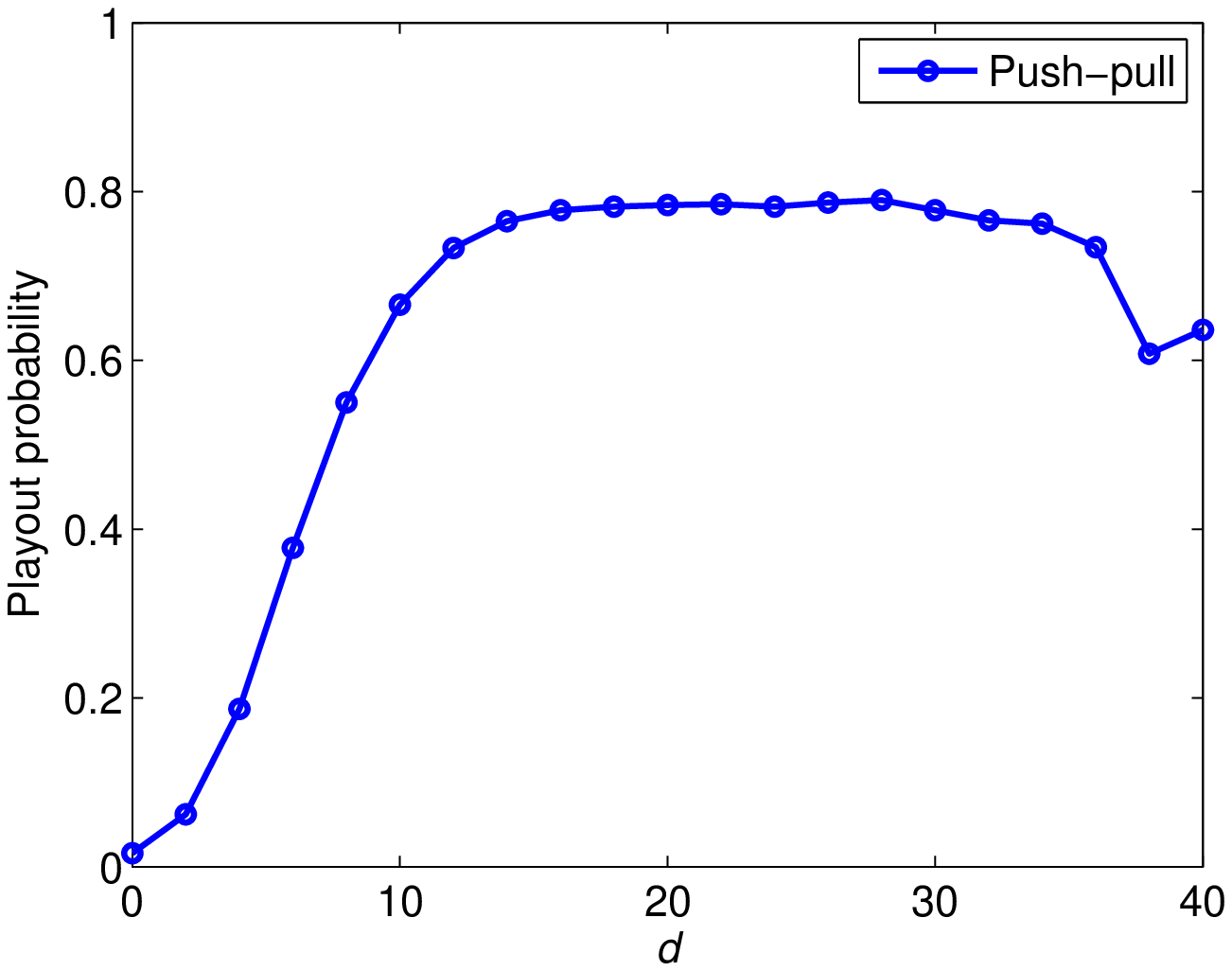}}
\subfigure[]{
\label{fig:subfig:b}
\includegraphics[width=0.4\textwidth]{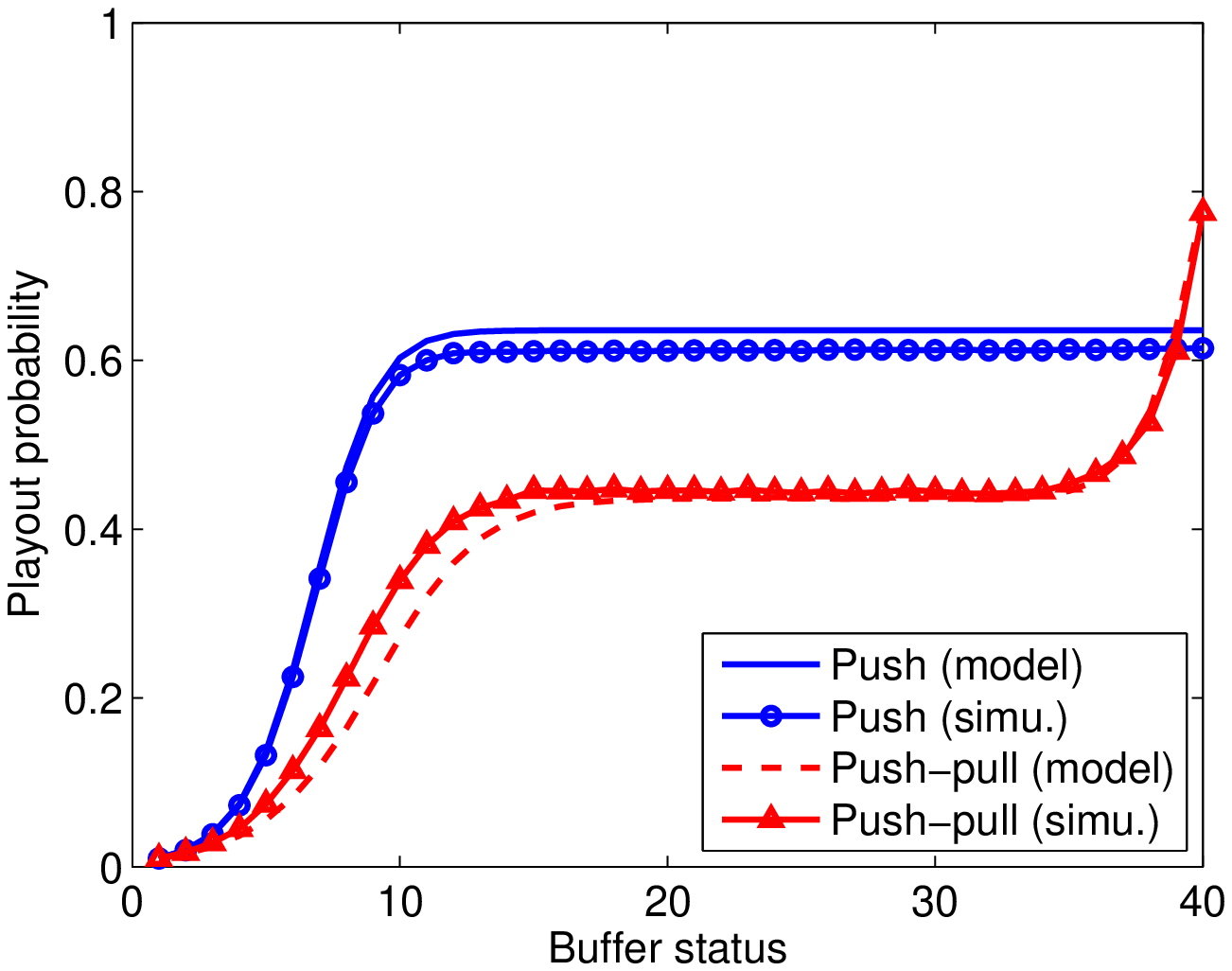}}
\caption{(a) Playout probability vs. $d$; (b) Buffer status of push scheme and push-pull scheme ($d=20$).}
\label{fig:pc}
\end{figure}
\section{Conclusions}
In this paper, we have made a detailed study on modeling and performance analysis of the pull-based P2P streaming systems. We establish the analytical framework for the pull-based streaming schemes in P2P network, give accurate models of the chunk selection and peer selection strategies, and then organize them into three categories of streaming schemes, i.e., chunk first scheme, peer first scheme and epidemic scheme. The former two schemes need all peers' buffer maps information, while the epidemic scheme not.

\begin{itemize}
  \item In all of our considered cases, the performance of chunk first scheme is more sensitive to the increase of neighbor set compared to that in the peer first scheme. The playout probability decreases considerably with the increasing of the neighbor set size.
  \item Pull-based schemes do not perform as well as the push-based schemes when peers are limited to reply only one request in each time slot. Through both numerical analysis and simulation, we find all pull-based streaming schemes will reach close to optimal playout probability when the reply number is larger than three, especially for the peer first scheme.
  \item Although very simple, the random strategy is a fairly good chunk selection strategy. In the chunk first scheme, it is superior to the well-known latest first strategy and the greedy strategy in terms of the playout probability.
  \item As regard to the epidemic scheme, we demonstrate that the pure pull scheme has particularly poor performance than the existent push-based scheme. To this end, we propose a push-pull scheme which can considerably improve the playout probability without any buffer map information exchanged as well.
\end{itemize}

This paper throws lights on some fundamental problems in pull-based P2P streaming systems. However, there are also some stimulating and challenging directions to extend this work. First, the impacts of some important factors in the overlay topology deserve further research [23], e.g., the degree distribution, the clustering coefficient, etc., especially when peers have heterogeneous bandwidth. Second, how to model and improve the streaming schemes in combination with the incentive mechanism existent in many P2P systems will be an innovative subject worth probing into.

\section*{Acknowledgement}
This paper is supported by the National Key Technology R\&D Program (No. 2012BAH03F00).

\end{document}